\documentclass[trackchanges]{aastex7}

\shorttitle{Methyl Carbamate}
\shortauthors{Duan et al.}

\begin{document}

\title{First Interstellar Detection of Methyl Carbamate: A New Observational Anchor for Glycine Chemistry}

\correspondingauthor{Qian Gou}
\email{qian.gou@cqu.edu.cn}

\author[orcid=0009-0002-5477-6309]{Chunguo Duan}
\affiliation{School of Chemistry and Chemical Engineering, Chongqing University, Daxuecheng South Rd. 55, Chongqing 401331, People’s Republic of China}
\affiliation{Chongqing Key Laboratory of Chemical Theory and Mechanism, Daxuecheng South Rd. 55, Chongqing 401331, People’s Republic of China}
\email{cquduancg@cqu.edu.cn}  

\author[orcid=0000-0001-5950-1932]{Fengwei Xu} 
\affiliation{Max Planck Institute for Astronomy, Königstuhl 17, Heidelberg 69117, Germany}
\email{fengweilookuper@gmail.com}

\author[orcid=0009-0004-1066-4924]{Jun Kang} 
\affiliation{School of Chemistry and Chemical Engineering, Chongqing University, Daxuecheng South Rd. 55, Chongqing 401331, People’s Republic of China}
\affiliation{Chongqing Key Laboratory of Chemical Theory and Mechanism, Daxuecheng South Rd. 55, Chongqing 401331, People’s Republic of China}
\email{jun.kang@stu.cqu.edu.cn}

\author[orcid=0000-0003-3831-3582]{Qian Gou} 
\affiliation{School of Chemistry and Chemical Engineering, Chongqing University, Daxuecheng South Rd. 55, Chongqing 401331, People’s Republic of China}
\affiliation{Chongqing Key Laboratory of Chemical Theory and Mechanism, Daxuecheng South Rd. 55, Chongqing 401331, People’s Republic of China}
\email{qian.gou@cqu.edu.cn}

\author[orcid=0000-0001-8514-6989]{Xuefang Xu} 
\affiliation{School of Chemistry and Chemical Engineering, Chongqing University, Daxuecheng South Rd. 55, Chongqing 401331, People’s Republic of China}
\affiliation{Chongqing Key Laboratory of Chemical Theory and Mechanism, Daxuecheng South Rd. 55, Chongqing 401331, People’s Republic of China}
\email{xuefang_xu@cqu.edu.cn}

\author[orcid=0000-0002-3319-1021]{Laurent Pagani} 
\affiliation{LUX, Observatoire de Paris, PSL Research University, CNRS, Sorbonne Universités, Paris 75014, France}
\email{laurent.pagani@obspm.fr}

\author[orcid=0009-0005-1132-5876]{Jiaxin Du} 
\affiliation{School of Chemistry and Chemical Engineering, Chongqing University, Daxuecheng South Rd. 55, Chongqing 401331, People’s Republic of China}
\affiliation{Chongqing Key Laboratory of Chemical Theory and Mechanism, Daxuecheng South Rd. 55, Chongqing 401331, People’s Republic of China}
\email{jiaxindu@stu.cqu.edu.cn}

\author[orcid=0000-0002-5435-925X]{Xi Chen} 
\affiliation{Center for Astrophysics, Guangzhou University, Guangzhou 510006, People’s Republic of China}
\email{chenxi@gzhu.edu.cn}

\begin{abstract}

Glycine—the simplest amino acid—has remained undetected in the interstellar medium despite decades of sensitive searches, motivating alternative approaches to constrain its astrochemical origin. A promising strategy is to investigate the broader $\rm C_{2}H_{5}O_{2}N$ isomer family and identify detectable members that can serve as observational anchors for glycine-related chemistry. Herein, we report the first robust interstellar detection of methyl carbamate toward the hot molecular core G358.93-0.03 MM1 using ALMA 1 mm observations. Ten unblended rotational transitions are identified, yielding a column density of (4.21$\pm0.84)\times10^{15} \rm cm^{-2}$ and an excitation temperature of $204\pm10$ K. We also searched for other $\rm C_{2}H_{5}O_{2}N$ isomers with available rotational spectroscopic data, including glycine, but none were detected, allowing us to derive upper limits on their column densities. The resulting abundance pattern deviates significantly from the Minimum Energy Principle predictions, highlighting that the $\rm C_{2}H_{5}O_{2}N$ family is shaped primarily by kinetic chemical process rather than thermodynamic equilibrium. The observed methyl carbamate abundance is consistent with a grain-surface formation scenario involving radical–radical recombination ($\rm CH_{3}$O + $\rm NH_{2}$CO), further supported by its correlated abundances with its proposed precursors, methanol and formamide, across diverse astrophysical environments. This detection establishes methyl carbamate as a new observational anchor for glycine chemistry, providing critical constraints on the formation pathways of amino-acid-related molecules in star-forming regions.

\end{abstract}


\section{Introduction} \label{sec:1}

The origin of molecular complexity relevant to the emergence of life is a central question in modern astrochemistry. Star-forming regions provide environments in which complex organic molecules (COMs) can form prior to planet formation, potentially seeding nascent planetary systems with prebiotic material \citep{1990Sci...249..366C, 1992Natur.355..125C, 2009ARA&A..47..427H}. Among such species, amino acids occupy a unique position due to their fundamental roles in terrestrial biochemistry. Yet, despite decades of increasingly sensitive searches, the simplest amino acid, glycine ($\rm NH_{2}CH_{2}C(O)OH$), has not been securely detected in the interstellar medium (ISM) \citep{2003ApJ...593..848K, 2005ApJ...619..914S, 2007MNRAS.376.1201C, 2007MNRAS.374..579J, 2016ApJ...830L...6J, 2020AsBio..20.1048J}, even though it is well established in meteorites, comets, and returned asteroid samples \citep{1983AdSpR...3i...5C, 2012cosp...39..264B, 2016SciA....2E0285A, 2023NatCo..14.1482P}. This long-standing non-detection suggests that the challenge is not merely observational, but reflects an incomplete understanding of the chemical pathways governing amino-acid-related molecules in space.

An indirect but informative approach to this problem is to investigate the broader $\rm C_{2}H_{5}O_{2}N$ isomer family, which includes glycine and several chemically related species with distinct functional groups, thermodynamic stabilities, and spectroscopic properties. Beyond expanding the search for glycine-related molecules, this approach provides a way to test whether the abundances of such species are controlled mainly by thermodynamic stability or by pathway-specific chemistry. The relative abundances of interstellar isomers are often discussed in the context of the Minimum Energy Principle \citep[MEP;][]{2009ApJ...696L.133L}, which predicts that, under thermodynamic control, the most stable (lowest-energy) member of an isomer set should dominate. In this framework, molecular abundances are expected to decline approximately as exp(-$\Delta E$/$T_{\rm kin}$) with increasing relative energy $\Delta E$ at kinetic temperature $T_{\rm kin}$ \citep{2021ApJ...912L...6G}. However, for constitutional isomers, which share the same elemental composition but differ in bonding connectivity, interstellar abundances do not always follow simple stability ordering. A well-known example is the $\rm C_{2}H_{4}O_{2}$ family, where methyl formate ($\rm CH_{3}OCHO$), glycolaldehyde ($\rm CH_{2}OHCHO$), and acetic acid ($\rm CH_{3}COOH$) exhibit abundance ratios that deviate from thermodynamic expectations, demonstrating that pathway-specific kinetics can dominate over equilibrium chemistry in warm star-forming regions \citep{2020A&A...644A..84M}.

Whether the same principle applies to glycine-related $\rm C_{2}H_{5}O_{2}N$ chemistry remains unclear. This uncertainty largely reflects the long-standing absence of secure interstellar detections of glycine and its chemically related isomers, which has prevented any empirical assessment of how this prebiotic family is populated in space \citep{2004sf2a.conf..493D, 2020AA...639A.135S, 2021A&A...653A.129C, 2022AA...666A.134S}. Recent detection of glycolamide ($\rm HOCH_{2}C(O)NH_{2}$, hereafter GA) toward the Galactic Center cloud G+0.693-0.027 \citep{2023ApJ...953L..20R} and the hot molecular core (HMC) G358.93-0.03 MM1 \citep[hereafter G358.93 MM1,] []{Duan26}, has provided the first observational foothold into $\rm C_{2}H_{5}O_{2}N$ chemistry. \citet{2023ApJ...953L..20R} suggested that this family is kinetically controlled in G+0.693-0.027, a relatively cool source with a kinetic temperature of $\sim$100 K \citep{2018MNRAS.478.2962Z}. However, whether this conclusion also holds in hotter environments such as HMCs remains unclear, because higher temperatures might in principle bring the chemistry closer to thermodynamic control. With only one confirmed member in G+0.693-0.027 and G358.93 MM1, additional detections are therefore needed to determine whether thermodynamic stability or pathway-specific kinetics dominates glycine-related chemistry across the diverse physical conditions.

Methyl carbamate ($\rm CH_{3}OC(O)NH_{2}$, hereafter MC) is pivotal in this context. As a member of the $\rm C_{2}H_{5}O_{2}N$ isomer family, MC occupies a particularly informative position: it is among the more stable isomers \citep{2019ESC.....3.1170S}, and is structurally distinct from both glycine and glycolamide. Importantly, the laboratory rotational spectrum of MC has been well characterized \citep{Mar1999, BAKRI2002312, ILYUSHIN2006127}, and its large permanent dipole moment ($\sim$2.4 Debye) makes it a favorable target for millimeter-wave observations. Despite these favorable properties and repeated predictions from astrochemical models, MC has not previously been securely detected in the ISM \citep{2004sf2a.conf..493D, 2020ApJ...899...65S}. 

The absence of a secure detection of MC to date therefore raises a natural question of whether suitable physical conditions and observational targets have been available to test these expectations. Astrochemical models provide clear guidance in this respect: simulations predict that MC can form efficiently in warm, dense environments dominated by active grain-surface chemistry \citep[e.g.,][]{Gar13, 2020ApJ...899...65S, Gar22}. In particular, \citep{Gar22} showed that MC can reach abundances in HMCs comparable to that of GA. The HMC G358.93 MM1, where glycolamide has already been securely detected \citep{Duan26}, thus provides a well-motivated target in which to search for methyl carbamate.

Using deep ALMA 1 mm observations, we report the first robust detection of MC toward G358.93 MM1. This detection establishes MC as a new observational anchor for the $\rm C_{2}H_{5}O_{2}N$ isomer family, providing stringent constraints on isomer abundances and enabling a first empirical assessment of the chemical processes shaping glycine-related molecules in star-forming regions. The remainder of this Letter is organized as follows: Section \ref{sec:2} describes the observations and data reduction; Section \ref{sec:3} presents the spectral analysis and derived molecular parameters; Section \ref{sec:4} discusses the chemical implications; Section \ref{sec:5} summarizes the main conclusions.

\begin{figure}[]
\centering
\includegraphics[width=1\columnwidth]{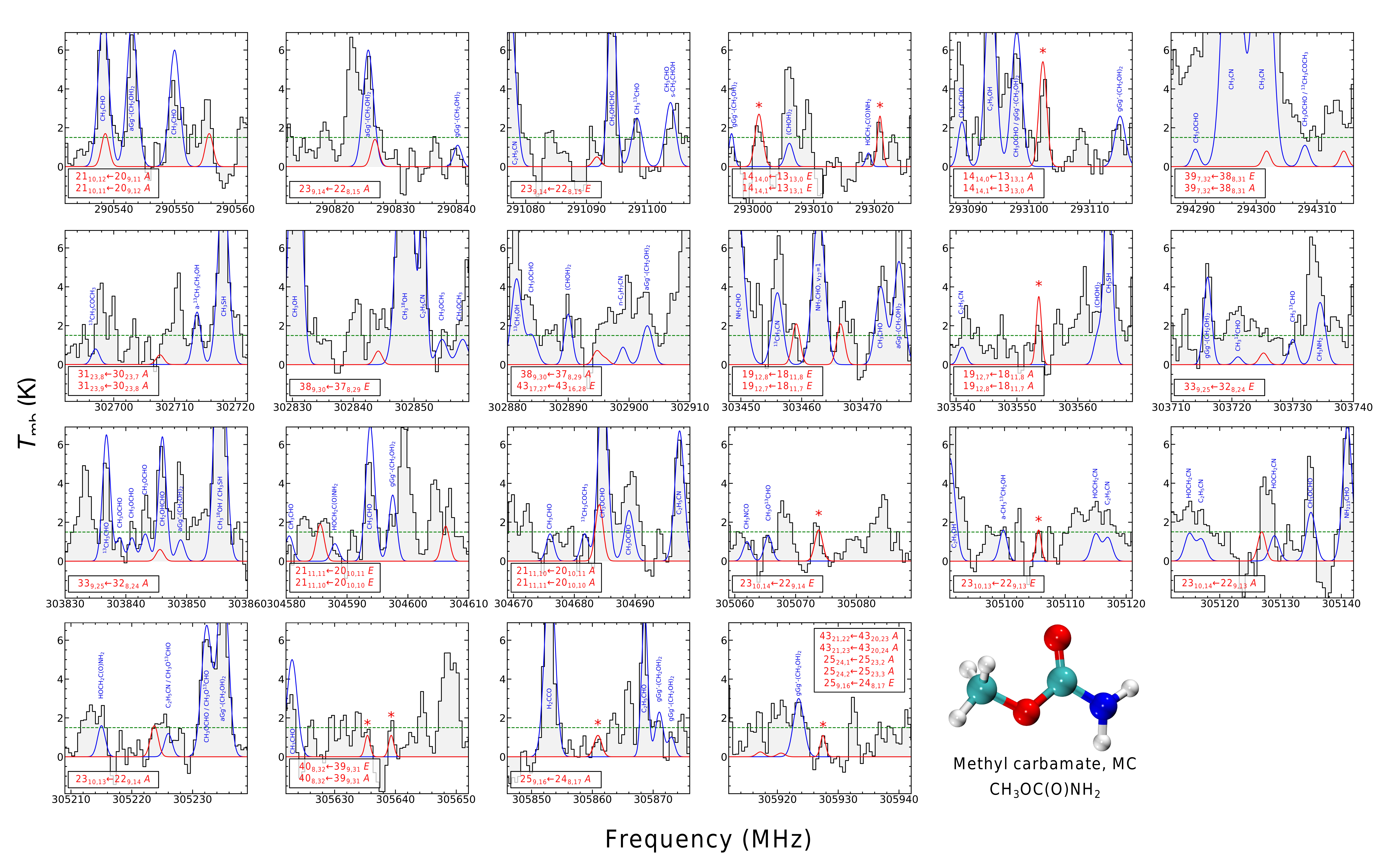}
\caption{MC emission toward G358.93 MM1. Black histograms show the spectra extracted at the offset position described in Appendix \ref{sec:appA}. Red curves represent the best-fit LTE model for MC, while blue curves indicate the full spectral model including all other species identified toward this source. Green dashed lines mark the 3$\sigma$ noise level. Red asterisks highlight the ten MC transitions that are unblended and used to derive the molecular parameters. The inset shows the molecular structure of MC (C: gray; O: red; N: blue; H: white).
\label{fig:1}}
\end{figure}

\section{Observation and data reduction} \label{sec:2}

The high-mass star-forming region G358.93-0.03 was observed with the Atacama Large Millimeter/submillimeter Array (ALMA) at $\sim$1 mm on 2021 May 17 (Project ID: 2019.1.00768.S; PI: Crystal Brogan). The phase center was set to R.A.(J2000) = $\rm 17^h 43^m 10^s$.000, Dec.(J2000) = $-29^\circ 51^\prime 46^{\prime \prime}$.000, with an on-source integration time of $\sim$49 minutes. The observations consisted four spectral windows (SPWs 1-4), each with a bandwidth of 1.875 GHz, centered at 291.4, 293.4, 303.4, and 305.1 GHz. The channel spacing is 488.24 kHz, corresponding to a velocity resolution ($\delta V$) of 0.5 km $\rm s^{-1}$ at these frequencies. The resulting data achieve an angular resolution of $\sim$$0.12^{\prime \prime}$ and a typical rms noise level of 0.5 K per 0.5 km $\rm s^{-1}$ channel. The absolute flux calibration uncertainty is estimated to be $\sim$10\%. Additional details of the observational setup and data reduction are described in \citet{Duan26}.

\section{Results} \label{sec:3}

Spectroscopic predictions for MC were taken from the Jet Propulsion Laboratory catalog \citep[JPL\footnote{\url{https://spec.jpl.nasa.gov}},] []{Pic98}, based on the laboratory measurements of \citet{Mar1999, BAKRI2002312, ILYUSHIN2006127}. The ALMA spectra were modeled using the CLASS/Weeds \citep{Mar11} of Grenoble Image and Line Data Analysis Software (GILDAS\footnote{\url{http://www.iram.fr/IRAMFR/GILDAS}}) under the assumption of local thermodynamic equilibrium (LTE). The model includes the source size ($\theta$), line width ($\Delta V$), systemic velocity ($V_{\rm LSR}$), excitation temperature ($T_{\rm ex}$), and column density ($N_{\rm t}$). $V_{\rm LSR}$ were calibrated to 15.9 km s$^{-1}$ using the H$_{2}$CO line at 291948.067 MHz. The source size, $\theta$, was fixed to the deconvolved continuum angular size, and $\Delta V$ was determined from Gaussian fits to the line profiles, leaving $T_{\rm ex}$ and $N_{\rm t}$ as free parameters.

The MC integrated emission peaks close to the continuum maximum; however, spectra toward the core center are severely affected by line crowding and blending with other molecular species, which complicates unambiguous line identification and reliable LTE fitting. We therefore extracted the spectrum at a nearby offset position where the MC features are less confused and exhibit clean, near-Gaussian profiles. This position was selected not to maximize the absolute MC intensity, but to maximize the number of clean, unblended MC transitions available for robust identification and fitting. Given the line-rich spectrum of G358.93 MM1, we constructed a full source model including the molecules reported in previous studies and the species identified in this work, in order to assess potential blends and overlaps in the candidate MC lines. We identified ten unblended MC transitions toward G358.93 MM1, spanning upper-level energies $E_{\rm u}$ = 103-333 K. Seven of these transitions exceed a 3$\sigma$ peak-intensity threshold, where $\sigma$ denotes the rms noise measured in line-free channels. Additional tests using spectra extracted from multiple nearby positions are described in Appendix \ref{sec:appA}. Spectroscopic information for all unblended lines can be found in Appendix \ref{sec:appB}.

To refine the LTE results obtained from CLASS/Weeds, we optimized the initial model by CLASS/ADJUST using the Powell algorithm \citep{Powell64} implemented in the SciPy optimization library. Because the uncertainties are dominated not only by the fitting procedure, but also by line blending and the simplifying assumptions of the LTE model, we adopted a conservative uncertainty of 20\% for the derived $N_{\rm t}$ values. The best-fit LTE solution yields $N_{\rm t}$ = (4.21$\pm$0.84)$\times$10$^{15}$ cm$^{-2}$ and $T_{\rm ex}$ = 204$\pm$10 K, with the linewidth of MC set to 1.1 km s$^{-1}$. The derived molecular parameters are summarized in Table \ref{table:1}. Figure \ref{fig:1} compares the observed spectra with the best-fit MC LTE model, as well as with the full spectral model that accounts for emission from other identified species. 

Following the secure detection of MC and the recent detection of $syn$-GA \citep{Duan26}, we searched for other $\rm C_{2}H_{5}O_{2}N$ isomers with available rotational spectroscopic data, including glycine conformers I and II \citep[$\rm NH_{2}CH_{2}C(O)OH$,][]{1995ApJ...455L.201L, ILYUSHIN2006127}, $anti$-glycolamide \citep[$anti$-$\rm HOCH_{2}C(O)NH_{2}$,][]{Maris04, 2020AA...639A.135S}, and the $Z$ and $E$ isomers of acetohydroxamic acid \citep[$\rm CH_{3}C(O)NHOH$,][]{2022AA...666A.134S}. None of these species were detected, and 3$\sigma$ upper limits on their column densities were derived under the assumption of LTE and are reported in Table \ref{table:1}.

\begin{deluxetable}{cccccccc}
\tablenum{1}
\tablecaption{Molecular parameters in G358.93 MM1. \label{table:1}}
\tablewidth{0pt}
\tabletypesize{\scriptsize}
\tablehead{
\colhead{Species} & \colhead{Formula} & \colhead{Structure} &
\colhead{Detected} & \colhead{$T_{\rm ex}$} & \colhead{$N_{t}$}  & $\chi(\rm H_{2})^{a}$ & \colhead{Ref}\\
\colhead{} & \colhead{} & \colhead{} & \colhead{} &
\colhead{(K)} & \colhead{(10$^{14} \rm cm^{-2}$)} & \colhead{(10$^{-11}$)} & \colhead{} 
}
\startdata
 methyl carbamate (MC)	            & $\rm CH_{3}OC(O)NH_{2}$	     & \raisebox{-.5\height}{\includegraphics[width=12mm, height=8mm]{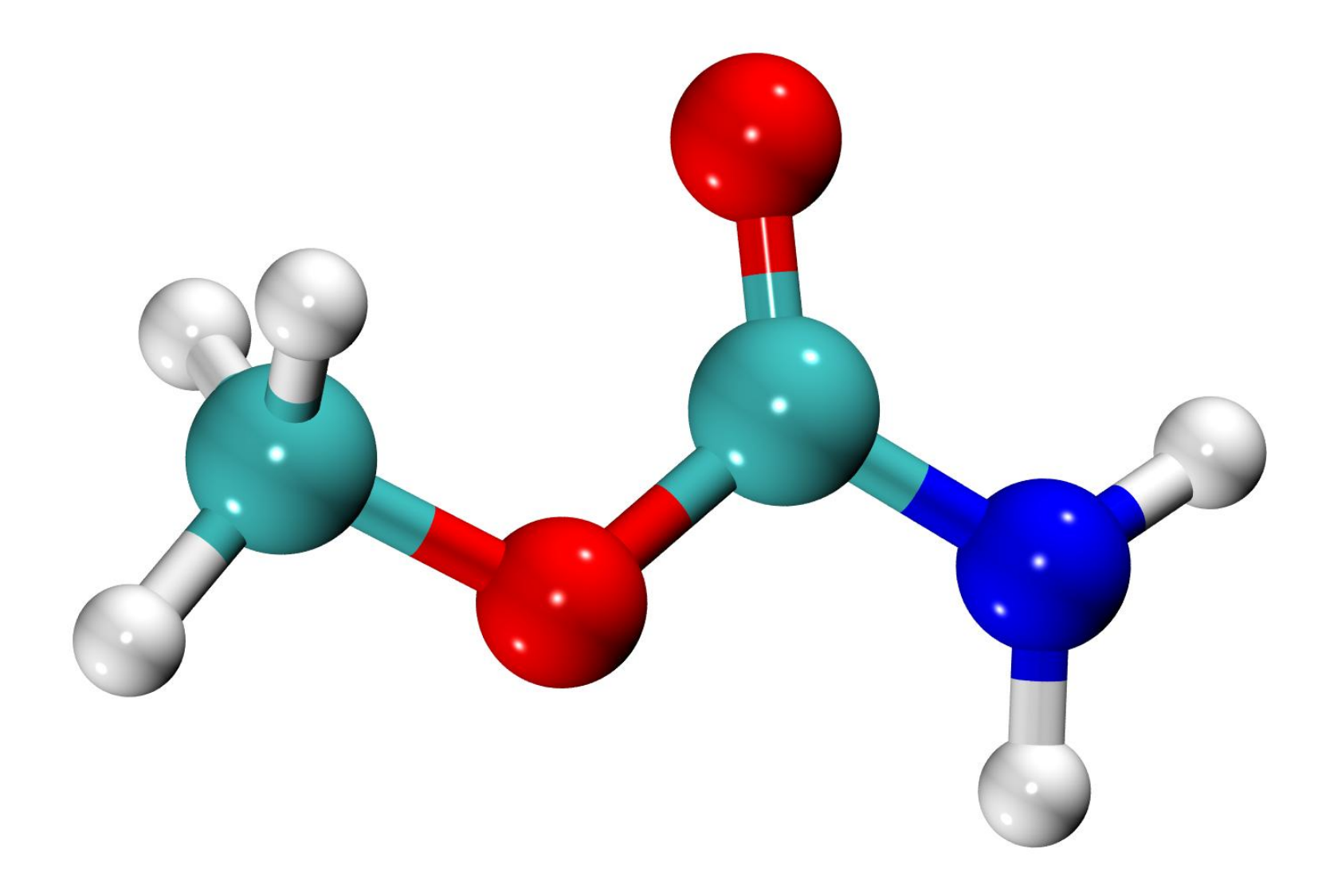}} & $\surd$   & 204$\pm$10	 & 42.1$\pm$8.4  & 53.8$\pm$15.2  & This work         \\ 
 glycine-I (GC-I)	                  & $\rm NH_{2}CH_{2}COOH$	     & \raisebox{-.5\height}{\includegraphics[width=12mm, height=8mm]{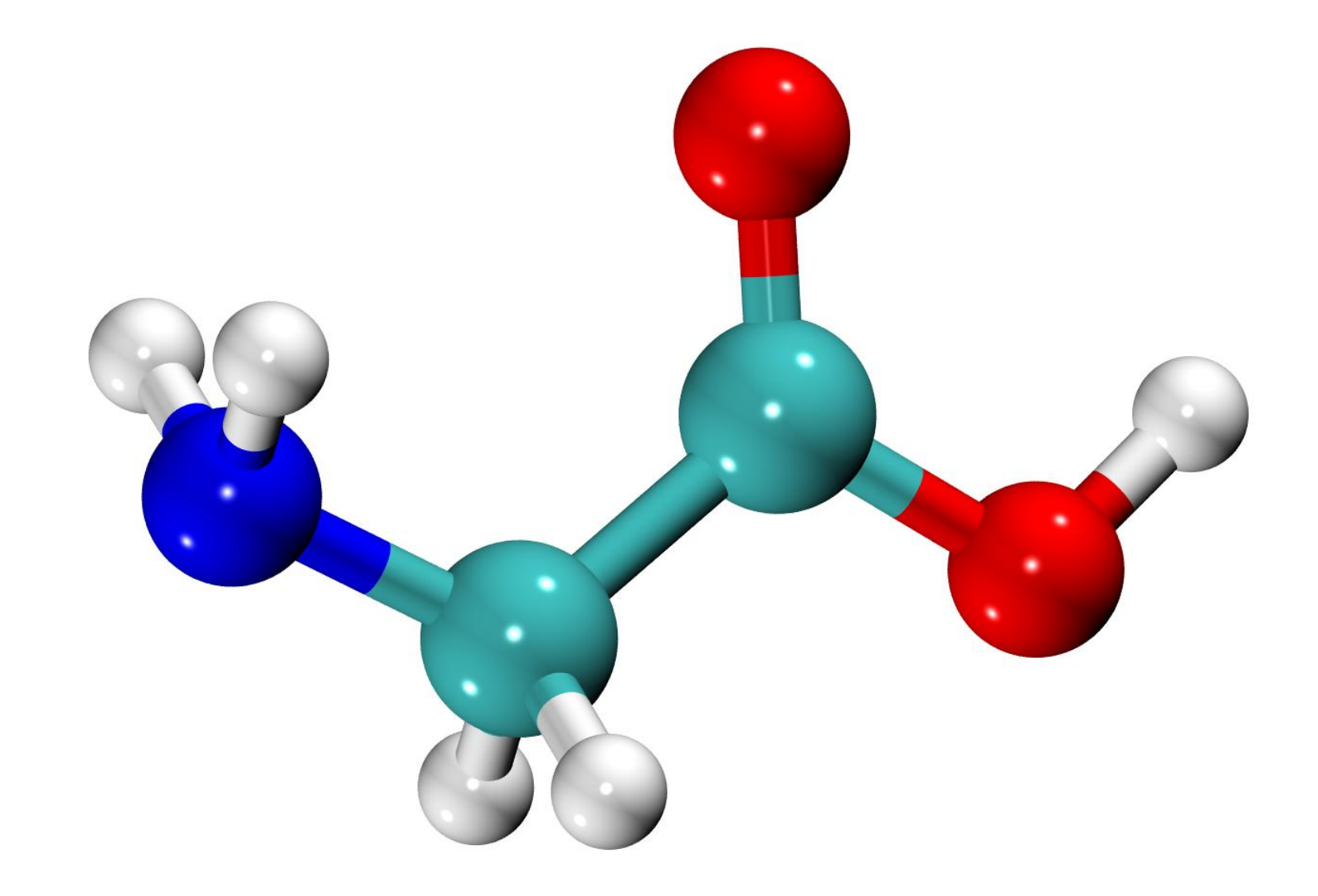}} & $\times$  &    -	       & $\leq$21.7$^{\rm b}$	 & $\leq$27.7$^{\rm b}$	 & This work         \\ 
 glycine-II (GC-II)	                & $\rm NH_{2}CH_{2}COOH$	     & \raisebox{-.5\height}{\includegraphics[width=12mm, height=8mm]{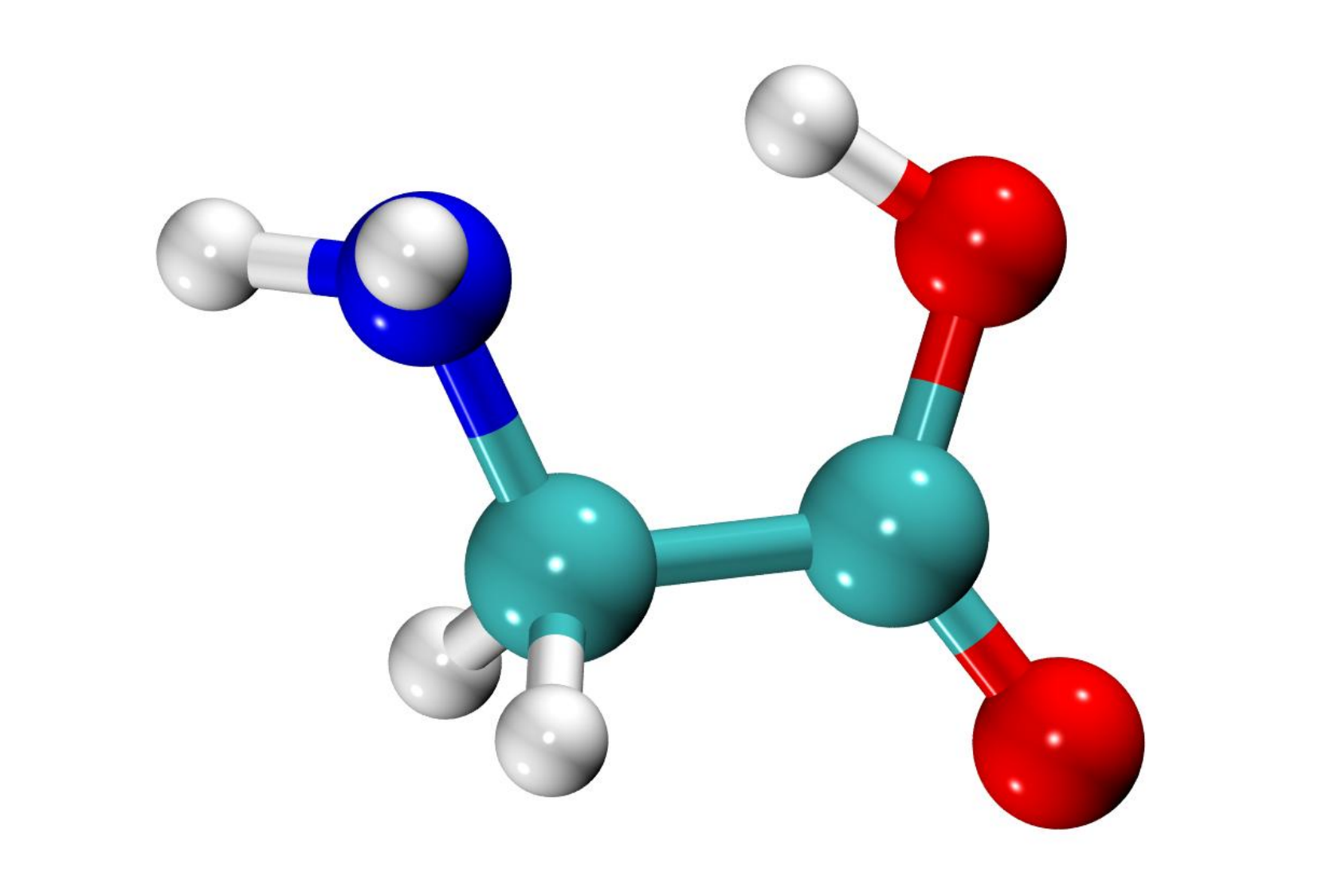}} & $\times$  &    -	       & $\leq$0.8$^{\rm b}$	   & $\leq$1.0$^{\rm b}$	   & This work         \\ 
 $syn$-glycolamide ($syn$-GA)	      & $\rm HOCH_{2}C(O)NH_{2}$     & \raisebox{-.5\height}{\includegraphics[width=12mm, height=8mm]{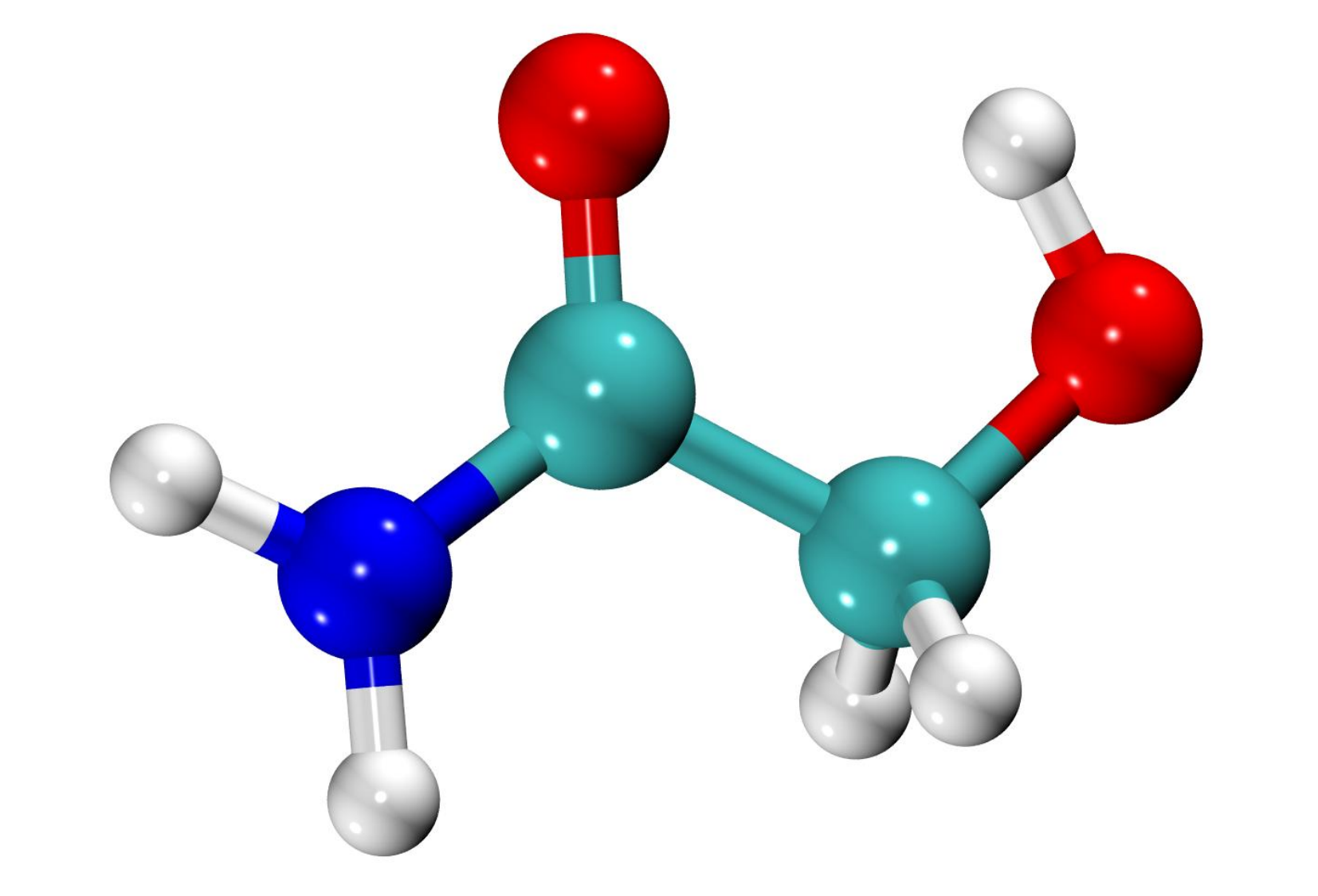}} & $\surd$   & 102$\pm$20$^{\rm c}$  & 5.3$\pm$1.1$^{\rm c}$  & 6.8$\pm$2.0$^{\rm c}$	 & \citet{Duan26}$^{\rm c}$  \\ 
 $anti$-glycolamide ($anti$-GA)	    & $\rm HOCH_{2}C(O)NH_{2}$     & \raisebox{-.5\height}{\includegraphics[width=12mm, height=8mm]{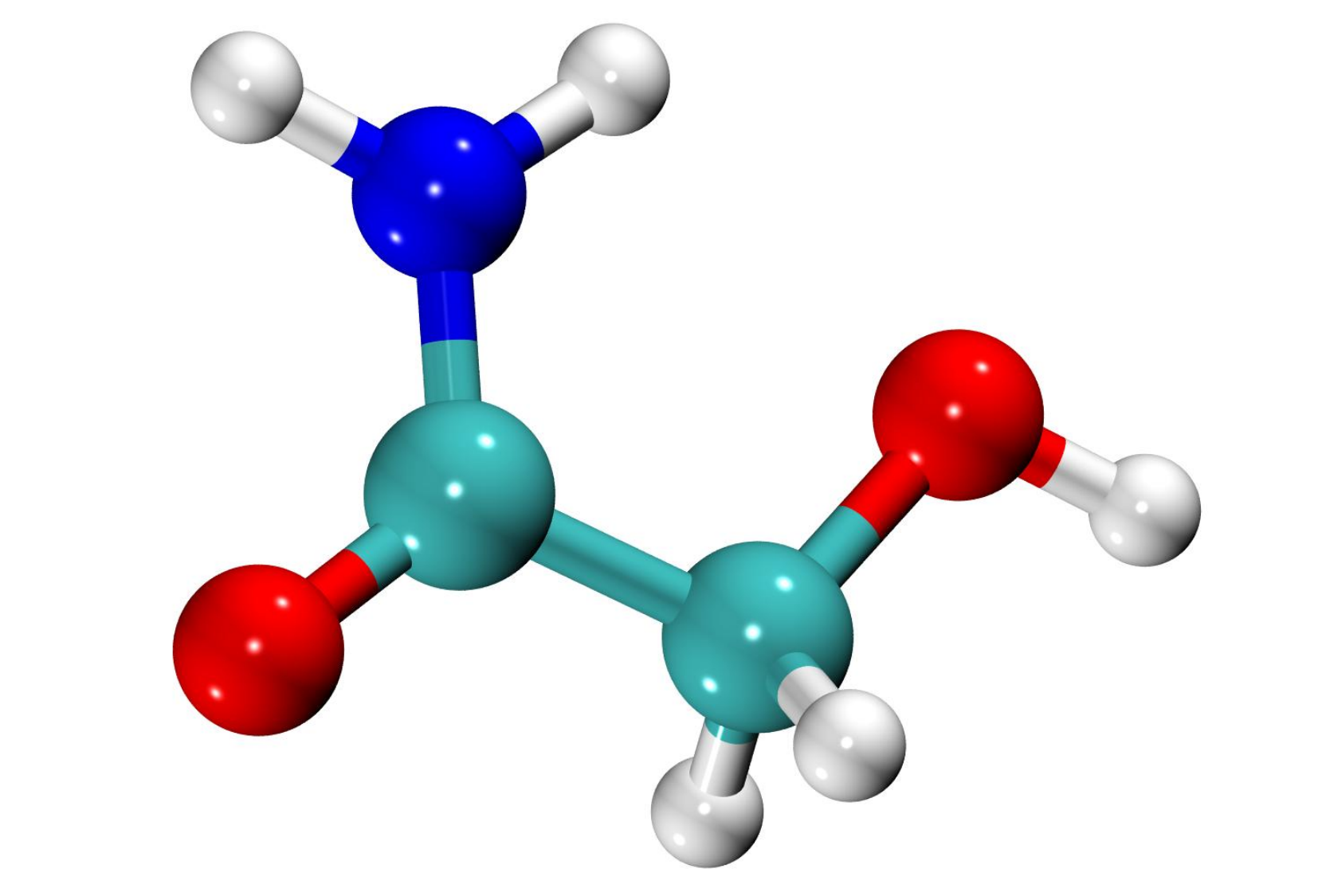}} & $\times$  &    -	       & $\leq$5.0$^{\rm b}$	   & $\leq$6.4$^{\rm b}$	   & This work         \\ 
 $Z$-acetohydroxamic acid ($Z$-AHA) & $\rm CH_{3}C(O)NHOH$	       & \raisebox{-.5\height}{\includegraphics[width=12mm, height=8mm]{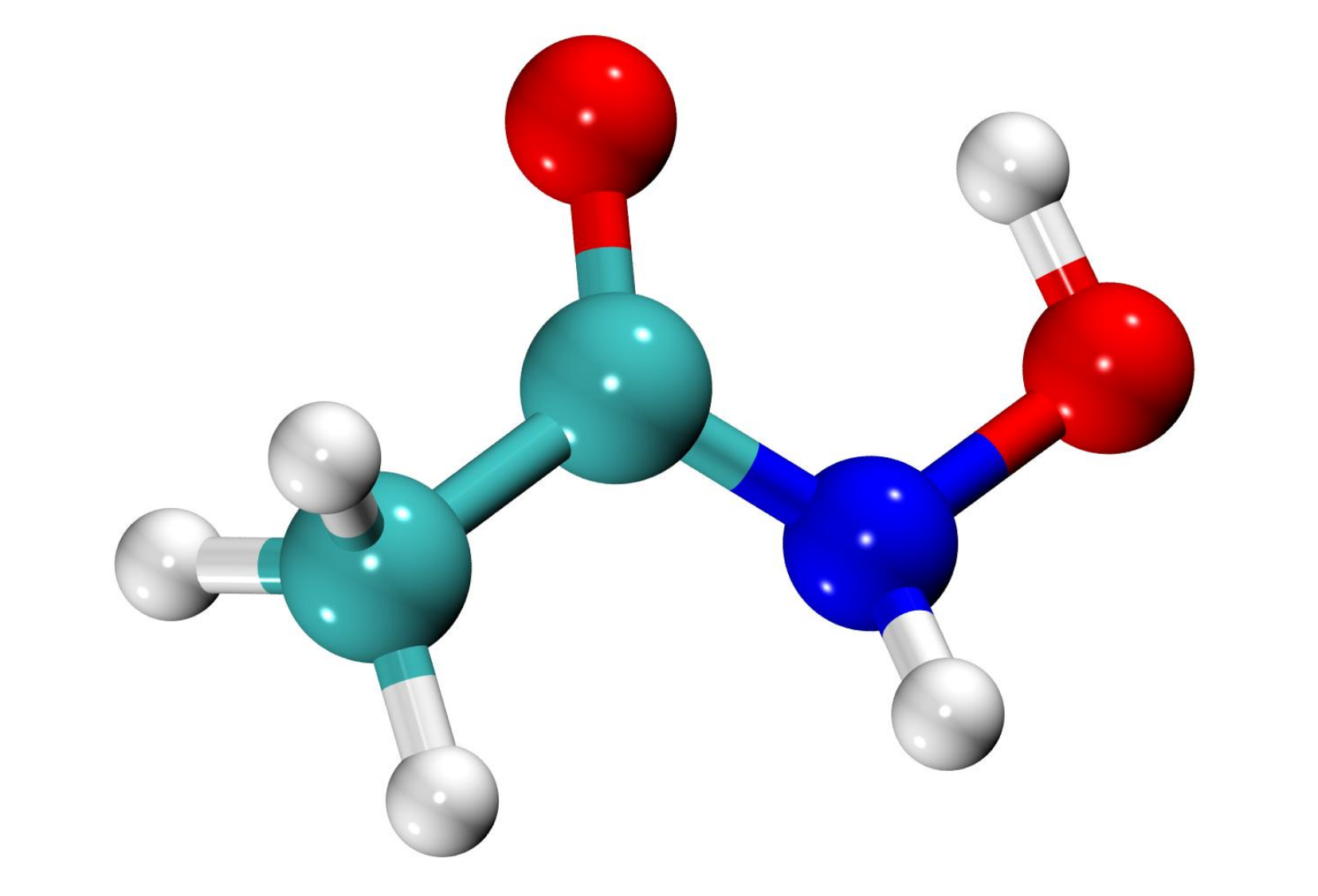}} & $\times$  &    -	       & $\leq$25.3$^{\rm b}$	 & $\leq$32.4$^{\rm b}$	 & This work         \\ 
 $E$-acetohydroxamic acid ($E$-AHA) & $\rm CH_{3}C(O)NHOH$	       & \raisebox{-.5\height}{\includegraphics[width=12mm, height=8mm]{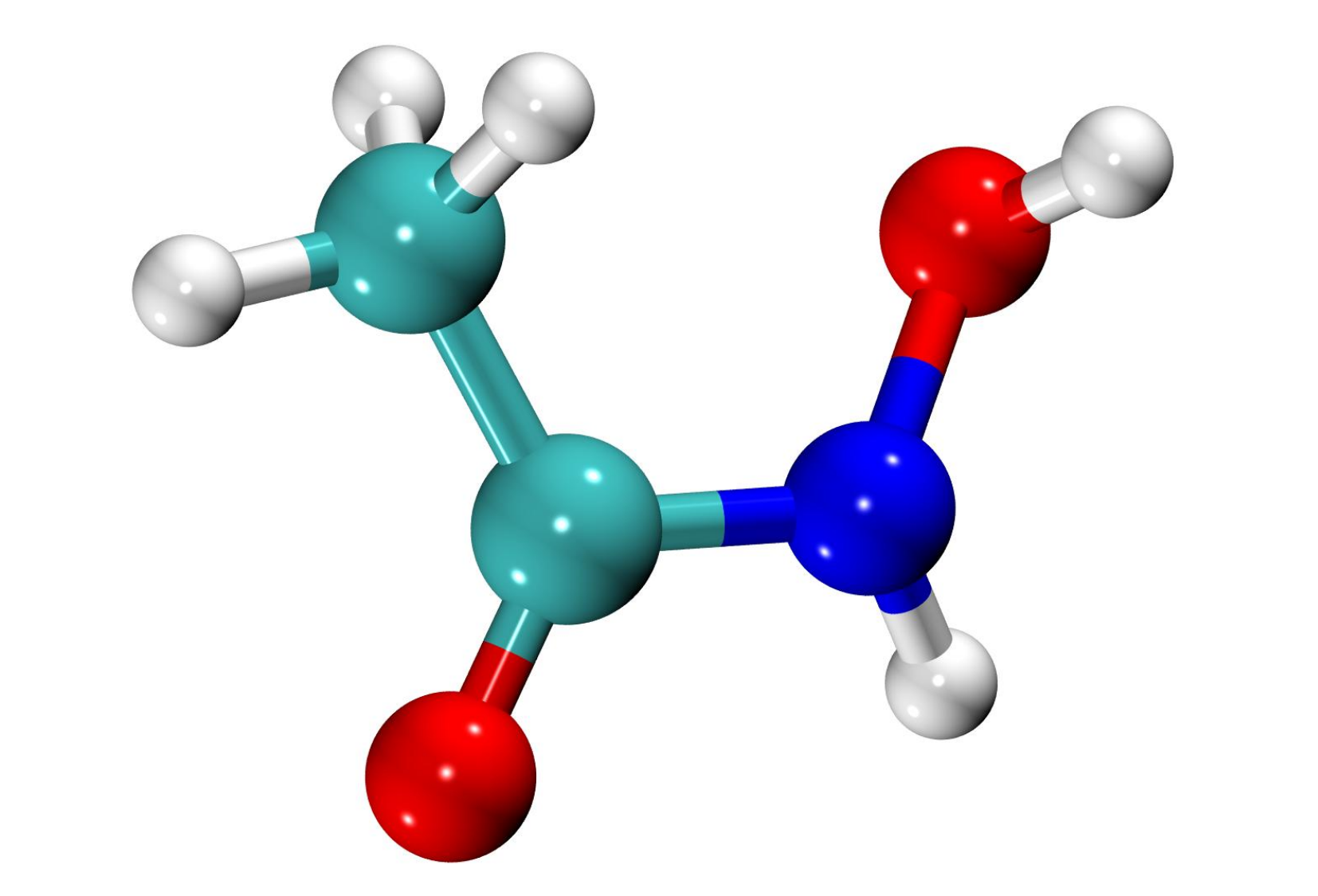}} & $\times$  &    -	       & $\leq$0.2$^{\rm b}$	   & $\leq$0.2$^{\rm b}$	   & This work         \\ 
\enddata
\tablecomments{
$^{\rm a}$$\chi(\rm H_{2})$ is the abundance relative to H$_{2}$, reported in units of $10^{-11}$. The adopted $N_{\rm H_{2}}$ and its derivation are described in Appendix \ref{sec:appC}.
$^{\rm b}$3$\sigma$ upper limits under the assumption of LTE.
$^{\rm c}$Parameters for $syn$-glycolamide ($syn$-GA) are taken from \citet{Duan26}}
\end{deluxetable}

\section{Discussion} \label{sec:4}

\subsection{Detectability of several key isomers in the $\rm C_{2}H_{5}O_{2}N$ family} \label{sec:4.1}

Despite sustained observational efforts targeting the $\rm C_{2}H_{5}O_{2}N$ family \citep[e.g.,][]{2004sf2a.conf..493D, 2020ApJ...899...65S, 2020AA...639A.135S, 2021A&A...653A.129C, 2022AA...666A.134S}, secure interstellar detections have remained rare. To date, only two members of this family have been firmly identified in the ISM: $syn$-GA \citep{2023ApJ...953L..20R, Duan26} and MC (this work). The emergence of these two species as the current observational footholds of the $\rm C_{2}H_{5}O_{2}N$ family raises the question of why they are detectable while other isomers remain elusive.

To assess whether spectroscopic factors alone can explain the observed detectability, we computed permanent dipole moments for several key $\rm C_{2}H_{5}O_{2}N$ isomers at the B2PLYP-D3(BJ)/aug-cc-pVTZ level (see Appendix \ref{sec:appD}). Among the species considered, glycine conformer II (GC-II) has the largest dipole moment ($\sim$5.5 Debye), followed by $syn$-glycolamide ($syn$-GA, $\sim$4.4 Debye), whereas MC has a more moderate value ($\sim$2.4 Debye). The fact that these higher-$\mu$ species remain undetected while MC and $syn$-GA are robustly detected demonstrates that spectroscopic favorability alone cannot account for the observed pattern, pointing instead to chemical abundance differences as the dominant factor.

\begin{figure}[]
\centering
\includegraphics[width=0.8\columnwidth]{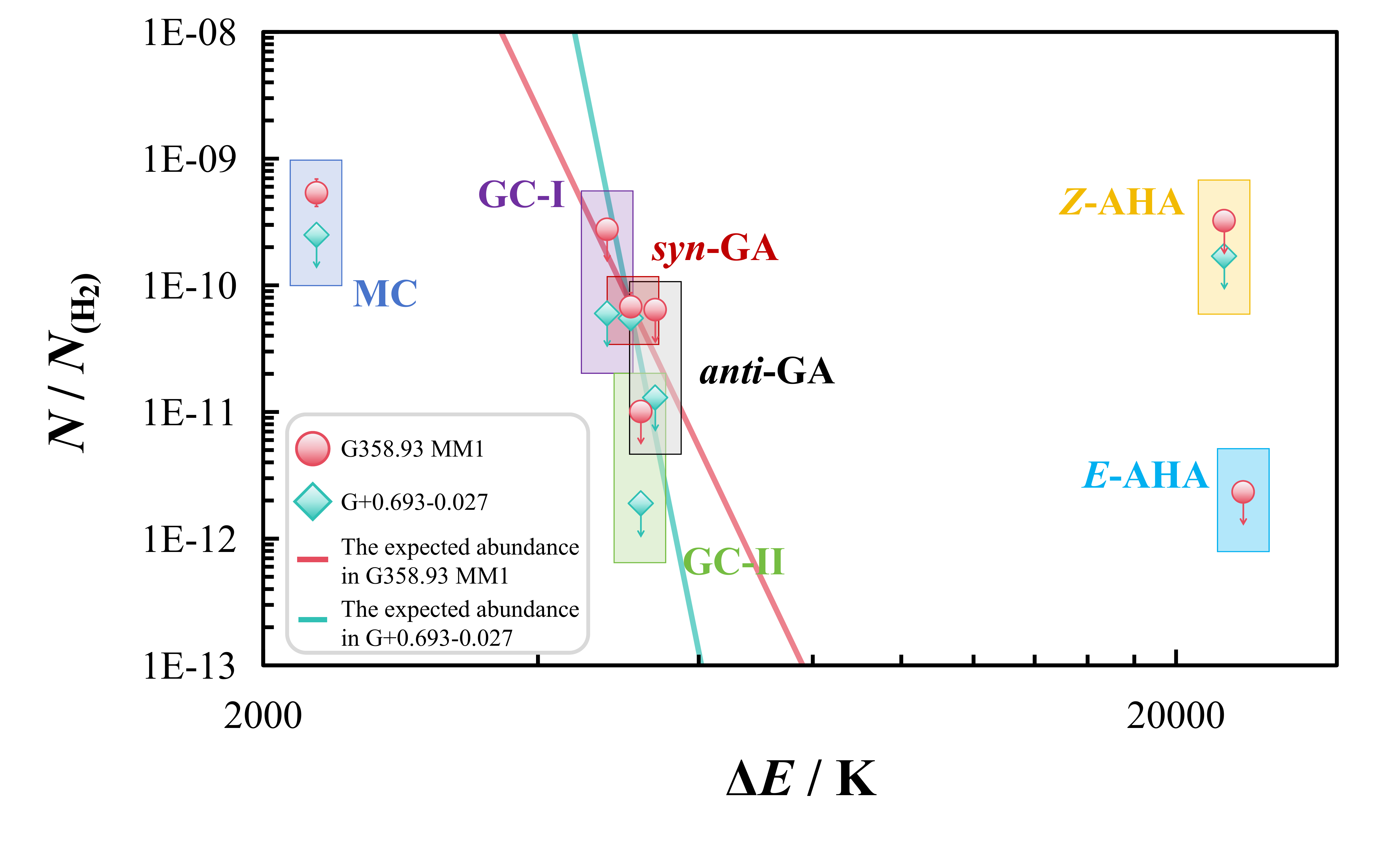}
\caption{Abundances relative to H$_{2}$ for C$_{2}$H$_{5}$O$_{2}$N isomers as a function of relative energy ($\Delta E$). Symbols denote measurements toward G358.93 MM1 and G+0.693-0.027, while arrows indicate 3$\sigma$ upper limits. The red and cyan lines show the abundance trends predicted from the MEP for G358.93 MM1 and G+0.693-0.027, respectively, constructed using representative kinetic temperatures and adopting $syn$-GA as a reference. For G358.93 MM1, $T_{\rm kin}$ = 172 K was adopted from CH$_{3}$CN \citep{2019ApJ...881L..39B}, while for G+0.693-0.027, $T_{\rm kin}$ = 100 K was adopted from \citet{2018MNRAS.478.2962Z}. Abundances for G+0.693-0.027 are taken from \citet{2023ApJ...953L..20R}. Labels indicate MC as methyl carbamate (CH$_{3}$OC(O)NH$_{2}$), while GC, GA, and AHA represent glycine (NH$_{2}$CH$_{2}$C(O)OH), glycolamide (HOCH$_{2}$C(O)NH$_{2}$), and acetohydroxamic acid (CH$_{3}$C(O)NHOH), respectively.
\label{fig:2}}
\end{figure}

Figure \ref{fig:2} places these observational results into a broader chemical context by comparing evaluated abundances and upper limits for the $\rm C_{2}H_{5}O_{2}N$ family with expectations based on thermodynamic stability. The observed abundance patterns in both G358.93 MM1 and G+0.693-0.027 deviate strongly from the trends expected under thermodynamic equilibrium, even when evaluated using $syn$-GA as the reference species and representative kinetic temperatures appropriate for each source. In G358.93 MM1, although MC is among the more stable detected members of the family and is robustly observed, its abundance lies well below the thermodynamic extrapolation. At the same time, several less stable but spectroscopically favorable isomers remain undetected. A similar departure from thermodynamic expectations is evident in G+0.693-0.027, where upper limits on multiple $\rm C_{2}H_{5}O_{2}N$ isomers further underscore the lack of correspondence between stability ordering and interstellar abundances. 

Because the LTE analysis was performed at an offset position rather than exactly at the continuum peak, we examined whether this choice could influence the kinetic-versus thermodynamic interpretation. The adopted position was selected for spectral reliability: the MC moment-0 maps (Appendix \ref{sec:appA}) show that MC emission is nearly co-spatial with the 1 mm continuum peak, but the continuum-peak spectra are too crowded for reliable fitting of individual MC transitions. In contrast, the offset spectrum provides the largest number of clean MC transitions. Although the warmer and denser gas near the continuum peak might in principle be more favorable to thermodynamic control, equilibrium among constitutional isomers would require efficient interconversion pathways on the relevant timescales, not simply higher temperature. Representative spectra extracted from nearby positions, including positions closer to the continuum peak, yield MC column densities within one order of magnitude of the adopted value. Such variations do not remove the discrepancy between the observed $\rm C_{2}H_{5}O_{2}N$ abundance pattern and the MEP-based expectation shown in Figure \ref{fig:2}. Therefore, the inferred kinetic behavior is unlikely to be an artifact of the adopted extraction position.

Taken together, the systematic deviation from thermodynamic trends across two chemically distinct environments shows that the $\rm C_{2}H_{5}O_{2}N$ family is not governed by equilibrium chemistry. Instead, the observed abundance pattern more likely reflects pathway-specific formation and destruction processes, with the detection of MC providing a critical new constraint on its interstellar formation. This interpretation is consistent with the behavior of other interstellar constitutional isomer families, such as $\rm C_{2}H_{4}O_{2}$, which also shows deviations from thermodynamic ordering in warm star-forming regions \citep{2020A&A...644A..84M}. Thus, our findings reinforce and extend the emerging picture that pathway-specific kinetics, rather than thermodynamic equilibrium, governs the interstellar abundances of the $\rm C_{2}H_{5}O_{2}N$ family in star-forming regions.

The stringent upper limits derived for glycine conformers I and II in G358.93 MM1 ($\leq$2.8$\times$10$^{-10}$ and $\leq$1.0$\times$10$^{-11}$, respectively; Table 1) suggest that glycine is at least a factor of $\sim$2 and $\sim$54 less abundant than MC in this source. This abundance deficit, together with the high evaporation temperature of glycine \citep[$\sim$200 K;][]{Gar13}, implies that any glycine emission is likely to be compact and susceptible to beam dilution. Detecting glycine in HMCs such as G358.93 MM1 will therefore require high spatial resolution observations, deep integrations, and sources with relatively narrow emission lines to reduce spectral confusion. The detection of MC (this work) and GA \citep{Duan26} demonstrates that multiple $\rm C_{2}H_{5}O_{2}N$ isomers are accessible with ALMA, suggesting that glycine may become detectable in carefully selected targets with sufficiently sensitive observations.

\subsection{Formation of methyl carbamate in space} \label{sec:4.2}

Motivated by the kinetically controlled behavior inferred for the $\rm C_{2}H_{5}O_{2}N$ family, we consider formation routes capable of producing MC efficiently in warm star-forming regions. High-temperature synthetic pathways known from terrestrial chemistry \citep[e.g., methanol + urea at temperatures $\geq$ 400 K,][]{sun2004} are unlikely to operate under typical interstellar conditions, pointing instead to grain-surface chemistry during the warm-up phase. Within this framework, MC can be formed through the grain-surface recombination of methoxy ($\rm CH_{3}O$) and carbamoyl ($\rm NH_{2}CO$) radicals, as proposed by \citet{2020ApJ...899...65S}:

\begin{equation}
\text{$\rm CH_{3}O$} + \text{$\rm NH_{2}CO$} \rightarrow \text{$\rm CH_{3}OC(O)NH_{2}$}
\label{eq:1}
\end{equation}
Quantum-chemical calculations at the B3LYP/6-311g++(d,p) level of theory indicate that Reaction \ref{eq:1} is strongly exothermic ($\sim$-80.8 kcal mol$^{-1}$) and likely barrierless \citep{2020ApJ...899...65S}, making it feasible on warm dust-grain surfaces where radicals become mobile. The $\rm CH_{3}O$ radical can be produced on grains via H-abstraction from methanol ($\rm CH_{3}OH$) or through recombination reactions involving $\rm CH_{3}$ and O \citep{Dean1987, 2020ApJ...899...65S, CJCP2304037}, while $\rm NH_{2}CO$ may arise from NH$_{2}$ + CO association, hydrogenation of HNCO, or H-abstraction from formamide \citep[$\rm NH_{2}CHO$,][]{Bel17, Bel19, Gar22}. The plausibility of this pathway in G358.93 MM1 is supported by the high abundances of the relevant precursor molecules. $\rm CH_{3}OH$ and $\rm NH_{2}CHO$ are both abundant in this source, yielding $\rm CH_{3}OH$/MC $\approx$ 513 (see Appendix \ref{sec:appD}) and $\rm NH_{2}CHO$/MC $\approx$ 4 (adopting the NH$_{2}$CHO column density from \citet{Duan26}), respectively. These ratios indicate ample reservoirs for the $\rm CH_{3}O$ and $\rm NH_{2}CO$ radicals during warm-up stage.

\begin{figure}[]
\centering
\includegraphics[width=1\columnwidth]{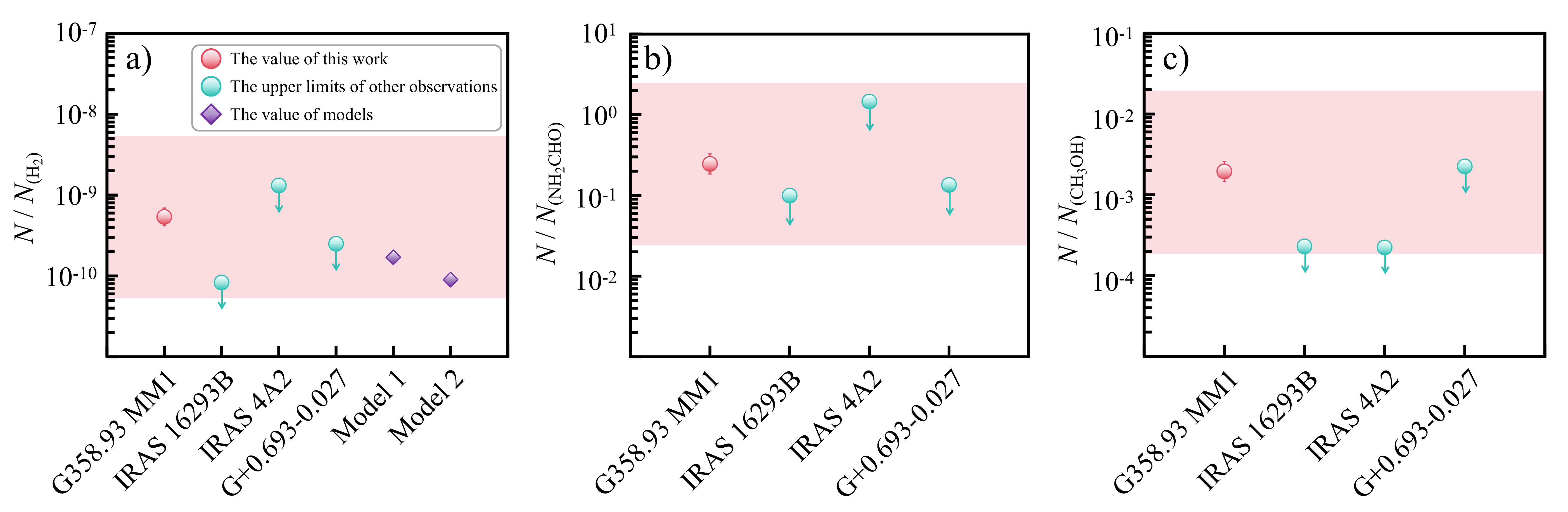}
\caption{Abundance constraints for methyl carbamate (MC) across different environments. a) MC abundance relative to H$_{2}$. b) MC abundance relative to NH$_{2}$CHO. c) MC abundance relative to CH$_{3}$OH. Red circles denote the values derived in this work toward the hot molecular core G358.93 MM1. Cyan circles represent upper limits from previous observations of hot corinos and the Galactic center cloud G+0.693-0.027, while purple diamonds indicate predictions from the astrochemical models of \citet{2020ApJ...899...65S}. The red shaded regions indicate the one order of magnitude uncertainty range associated with the MC abundance measured in this work, shown for reference across panels.
\label{fig:3}}
\end{figure}

Figure \ref{fig:3} places the MC abundance constraints into a broader context by comparing MC-to-precursor ratios (MC/$\rm CH_{3}OH$ and MC/$\rm NH_{2}CHO$) across environments spanning the hot molecular core G358.93 MM1 (this work), the hot corinos IRAS 16293B \citep{Cou16, 2020ARA&A..58..727J, 2020ApJ...899...65S} and NGC1333 IRAS 4A2 \citep[hereafter IRAS 4A2,][]{2017A&A...606A.121L, 2020ApJ...899...65S}, and the quiescent molecular cloud G+0.693-0.027 \citep{2023MNRAS.523.1448Z, 2023ApJ...953L..20R}. Despite the strong dependence of absolute COM abundances on temperature and evolutionary stage \citep{Gar13, Gar22}, these ratios are broadly consistent across sources within current uncertainties. This behavior suggests that MC production is chemically coupled, at least at the level of global scaling, to methanol and formamide that plausibly seed $\rm CH_{3}O$- and $\rm NH_{2}CO$-bearing radical chemistry.

Alternative formation routes have been proposed, such as $\rm NH_{2}$ addition to methoxycarbonyl \citep[$\rm CH_{3}OCO$,][]{2009ApJ...697..428L}, but the formation of $\rm CH_{3}OCO$ and its role in astrochemical networks remain uncertain. In this context, the grain-surface pathway in Reaction \ref{eq:1} provides a physically motivated baseline and is the dominant formation channel adopted in the models of \citet{2020ApJ...899...65S}. For G358.93 MM1, the measured MC abundance ($\sim$5.4$\times$10$^{-10}$) is consistent within an order of magnitude with its predicted value ($\sim$1.7$\times$10$^{-10}$; Figure \ref{fig:3}), supporting---but not uniquely confirming---a linked grain-surface formation scenario. We emphasize that this interpretation remains tentative, as only a single secure interstellar detection of MC is currently available. Further observational, laboratory, and theoretical studies will be required to establish its dominant formation pathways in space.

\section{Conclusions} \label{sec:5}

Using ALMA 1 mm observations, we present the first robust interstellar detection of methyl carbamate ($\rm CH_{3}OC(O)NH_{2}$, MC) toward the hot molecular core G358.93 MM1. From ten unblended transitions, we derived a column density of (4.21$\pm$0.84)$\times$10$^{15}$ cm$^{-2}$ and an abundance relative to $\rm H_{2}$ of (5.4$\pm$1.5)$\times$10$^{-10}$, establishing MC as the most abundant $\rm C_{2}H_{5}O_{2}N$ isomer detected in this source to date.

Together with the recent detection of glycolamide, this result provides the first observational footholds into the $\rm C_{2}H_{5}O_{2}N$ isomer family, enabling a direct empirical assessment of the chemical principles governing glycine-related molecules in the interstellar medium. The measured abundances and upper limits toward both G358.93 MM1 and G+0.693-0.027 deviate strongly from thermodynamic expectations, demonstrating that this family is not governed by equilibrium chemistry but instead reflects pathway-specific, kinetically controlled formation processes. Within this kinetic framework, our analysis supports grain-surface radical chemistry during warm-up as a plausible route for MC formation, particularly through the recombination of $\rm CH_{3}O$ and $\rm NH_{2}CO$ radicals. The observed abundance of MC and its ratios relative to methanol and formamide are broadly consistent with astrochemical model predictions across a range of environments, suggesting a chemically coupled origin. Nevertheless, we emphasize that these interpretations remain provisional, as only a single secure detection of MC is currently available.

More broadly, the detection of MC demonstrates that multiple members of the $\rm C_{2}H_{5}O_{2}N$ family can form efficiently in star-forming regions, highlighting the chemical richness accessible during the early stages of stellar evolution. By establishing MC as a new observational anchor for glycine-related chemistry, this work opens a pathway toward quantitatively testing amino-acid-related formation networks in space. Continued laboratory measurements, target observations, and chemical modeling will be essential to further constrain these pathways and their implications for the astrochemical origins of prebiotic molecules.

\begin{acknowledgments}

This work makes use of the following ALMA data: ADS/JAO.ALMA\#2019.1.00768.S. ALMA is a partnership of ESO (representing its member states), NSF (USA), and NINS (Japan), together with NRC (Canada), MOST and ASIAA (Taiwan, China), and KASI (Republic of Korea), in cooperation with the Republic of Chile. The Joint ALMA Observatory is operated by ESO, AUI/NRAO, and NAOJ. We are grateful for support from the National SKA Program of China (No. 2025SKA0120100), National Natural Science Foundation of China (Grant No. W2512014), Fundamental Research Funds for the Central Universities (Grant No. 2025CDJ-IAISYB-060), and Postdoctoral Fellows Excellence Support Program (Grant No. 2404013554893087).

\end{acknowledgments}

\begin{contribution}

Qian Gou supervised the project. Chunguo Duan developed the concept of the manuscript and conducted the data analysis and visualization. Fengwei Xu reduced the observational data. Jun Kang conducted Quantum-chemical calculations. Chunguo Duan, Fengwei Xu, Jun Kang, Qian Gou, Xuefang Xu, Laurent Pagani, Jiaxin Du and Xi Chen led the discussion on the interpretation of the results and writing of the manuscript. All authors contributed to the analysis or interpretation of the data and to the final version of the manuscript.

\end{contribution}

\facilities{ALMA}
\software{GILDAS ({\url{http://www.iram.fr/IRAMFR/GILDAS}})}

\clearpage

\appendix

\section{Spectrum Extraction Position And Spatial distribution of MC} \label{sec:appA}

\begin{figure*}[ht!]
\centering
\includegraphics[width=0.5\columnwidth]{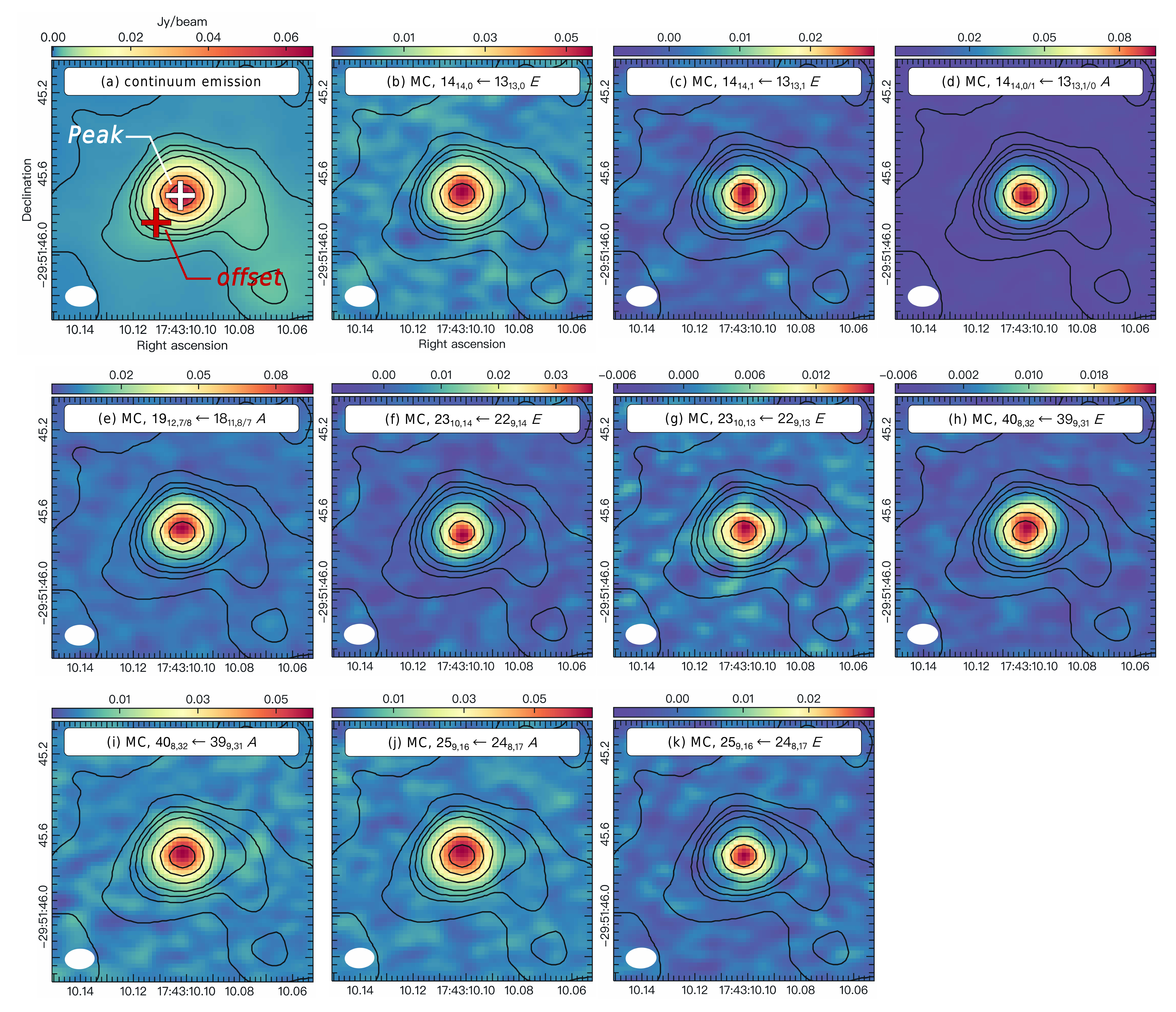}
\caption{(a) ALMA 1 mm continuum map of G358.93 MM1 and integrated intensity (Moment-0) maps of MC. Contours are shown at (20, 40, 80, 120, 200, 500, 1000, 1700)$\times \sigma$, where $\sigma$ is the rms noise measured in line-free regions. The synthesized beam (0.15$^{\prime \prime}\times0.10^{\prime \prime}$, position angle = -87.1$^{\circ}$) is presented as the white ellipse. The continuum peak is marked by the white cross, and the spectrum-extraction position adopted in this work is marked by the red cross. Each panel displays the upper level energy $E_{\rm u}$ above the map. (b)-(k) Integrated-intensity (moment-0) maps of MC toward G358.93 MM1, overlaid with the 1 mm continuum emission contours. The color scale unit is Jy beam$^{-1}$ km s$^{-1}$.
\label{fig:4}}
\end{figure*}

To mitigate absorption and severe line confusion at the continuum peak, we extracted spectra at a position offset from the 1 mm continuum maximum. The continuum peak itself shows strong MC emission, but the central spectra are severely affected by line crowding and blending with other molecular species, making unambiguous line identification and reliable LTE fitting difficult. The adopted offset position reduces spectral confusion and yields cleaner, approximately Gaussian profiles for many MC transitions. The extraction position used in this work is indicated in Figure \ref{fig:4}.

We examined spectra extracted from 70 offset positions around the continuum peak. MC emission is detectable at most of these positions, demonstrating that the detection is not confined to the single position adopted for LTE fitting. Among the tested positions, the adopted offset position provides the largest number of clean, ten in total, unblended MC transitions while maintaining a high signal-to-noise ratio. Although nearby positions generally provide fewer clean MC transitions because of stronger blending, representative LTE fits, including those closer to the continuum peak, yield MC column densities within one order of magnitude of the adopted value. Thus, although local abundance variations may exist, the secure detection of MC and the conclusion that the $\rm C_{2}H_{5}O_{2}N$ abundance pattern deviates from thermodynamic expectations are unlikely to be driven solely by the extraction position.

The spatial distribution of MC was analyzed using the Cube Analysis and Rendering Tool for Astronomy \citep[CARTA;][]{Com21}. Only transitions verified to be free of significant blending were used to generate the integrated-intensity (moment-0) maps. As shown in Figure \ref{fig:4}, the MC emission is nearly co-spatial with the 1 mm continuum peak of G358.93 MM1.

\clearpage

\section{Spectroscopic parameters of unblended MC transitions identified in G358.93 MM1} \label{sec:appB}

Table \ref{tab:2} lists spectroscopic parameters of all unblended or minimally blended transitions analyzed in this work, including rest frequencies, upper-state energies ($E_{\rm u}$), Einstein A-coefficients ($A_{\rm ij}$), degeneracies ($g_{\rm u}$), line width ($\Delta V$), and integrated intensity ($\int{T_{\rm MB}}dv$).

\begin{deluxetable*}{cccccccc}[htp]
\setlength{\tabcolsep}{13pt}
\tablenum{2}
\tablecaption{Spectroscopic parameters of unblended MC transitions identified in G358.93 MM1
\label{tab:2}}
\tablewidth{0pt}
\tablehead{
\colhead{Frequency} & \colhead{Upper level $\leftarrow$ Lower level$^{\rm a}$} & A/E$^{\rm b}$  &\colhead{$E_{\rm u}$} & \colhead{$A_{\rm ij}$} & \colhead{$g_{\rm u}$} & \colhead{$\Delta V$} & \colhead{$\int{T_{\rm MB}}dv$}  \\
\colhead{(MHz)} & \colhead{($J^{\prime}_{K_{a}^{\prime},K_{c}^{\prime}}$ $F^{\prime}$ $\leftarrow$ $J^{\prime\prime}_{K_{a}^{\prime\prime},K_{c}^{\prime\prime}}$ $F^{\prime\prime}$)} & \colhead{} &\colhead{(K)} & \colhead{($\rm s^{-1}$)} & \colhead{} & \colhead{(km $\rm s^{-1}$)} & \colhead{(K km $\rm s^{-1}$)} 
}
\startdata
293001.871  &  $14_{14,0}$ 13 $\leftarrow$ $13_{13,0}$ 12 &  E  &  103.4  &  7.17e-04  &  54  &  $1.57\pm0.27$  &  $3.53\pm0.32$  \\ 
293001.863  &  $14_{14,0}$ 14 $\leftarrow$ $13_{13,0}$ 13 &  E  &  103.4  &  7.18e-04  &  58  &      -          &      -          \\ 
293001.878  &  $14_{14,0}$ 15 $\leftarrow$ $13_{13,0}$ 14 &  E  &  103.4  &  7.21e-04  &  62  &      -          &      -          \\ 
293021.809  &  $14_{14,1}$ 13 $\leftarrow$ $13_{13,1}$ 12 &  E  &  103.3  &  7.18e-04  &  54  &  $1.01\pm0.26$  &  $2.93\pm0.26$  \\ 
293021.802  &  $14_{14,1}$ 14 $\leftarrow$ $13_{13,1}$ 13 &  E  &  103.3  &  7.18e-04  &  58  &      -          &      -          \\ 
293021.817  &  $14_{14,1}$ 15 $\leftarrow$ $13_{13,1}$ 14 &  E  &  103.3  &  7.22e-04  &  62  &      -          &      -          \\ 
293103.150  &  $14_{14,0}$ 13 $\leftarrow$ $13_{13,1}$ 12 &  A  &  103.4  &  7.18e-04  &  54  &  $1.38\pm0.06$  &  $6.78\pm0.31$  \\ 
293103.143  &  $14_{14,0}$ 14 $\leftarrow$ $13_{13,1}$ 13 &  A  &  103.4  &  7.19e-04  &  58  &      -          &      -          \\ 
293103.157  &  $14_{14,0}$ 15 $\leftarrow$ $13_{13,1}$ 14 &  A  &  103.4  &  7.22e-04  &  62  &      -          &      -          \\ 
293103.150  &  $14_{14,1}$ 13 $\leftarrow$ $13_{13,0}$ 12 &  A  &  103.4  &  7.18e-04  &  54  &      -          &      -          \\ 
293103.143  &  $14_{14,1}$ 14 $\leftarrow$ $13_{13,0}$ 13 &  A  &  103.4  &  7.19e-04  &  58  &      -          &      -          \\ 
293103.157  &  $14_{14,1}$ 15 $\leftarrow$ $13_{13,0}$ 14 &  A  &  103.4  &  7.22e-04  &  62  &      -          &      -          \\ 
303554.633  &  $19_{12,7}$ 18 $\leftarrow$ $18_{11,8}$ 17 &  A  &  117.4  &  5.03e-04  &  74  &  $1.08\pm0.12$  &  $3.20\pm0.28$  \\ 
303554.581  &  $19_{12,7}$ 19 $\leftarrow$ $18_{11,8}$ 18 &  A  &  117.4  &  5.03e-04  &  78  &      -          &      -          \\ 
303554.631  &  $19_{12,7}$ 20 $\leftarrow$ $18_{11,8}$ 19 &  A  &  117.4  &  5.04e-04  &  82  &      -          &      -          \\ 
303554.631  &  $19_{12,8}$ 18 $\leftarrow$ $18_{11,7}$ 17 &  A  &  117.4  &  5.03e-04  &  74  &      -          &      -          \\ 
303554.579  &  $19_{12,8}$ 19 $\leftarrow$ $18_{11,7}$ 18 &  A  &  117.4  &  5.03e-04  &  78  &      -          &      -          \\ 
303554.629  &  $19_{12,8}$ 20 $\leftarrow$ $18_{11,7}$ 19 &  A  &  117.4  &  5.04e-04  &  82  &      -          &      -          \\ 
305075.128  & $23_{10,14}$ 22 $\leftarrow$ $22_{9,14}$ 21 &  E  &  134.9  &  3.46e-04  &  90  &  $1.23\pm0.22$  &  $2.44\pm0.31$  \\ 
305075.118  & $23_{10,14}$ 23 $\leftarrow$ $22_{9,14}$ 22 &  E  &  134.9  &  3.46e-04  &  94  &      -          &      -          \\ 
305075.128  & $23_{10,14}$ 24 $\leftarrow$ $22_{9,14}$ 23 &  E  &  134.9  &  3.46e-04  &  98  &      -          &      -          \\ 
305106.610  & $23_{10,13}$ 22 $\leftarrow$ $22_{9,13}$ 21 &  E  &  134.9  &  3.46e-04  &  90  &  $0.88\pm0.19$  &  $1.71\pm0.13$  \\ 
305106.600  & $23_{10,13}$ 23 $\leftarrow$ $22_{9,13}$ 22 &  E  &  134.9  &  3.46e-04  &  94  &      -          &      -          \\ 
305106.609  & $23_{10,13}$ 24 $\leftarrow$ $22_{9,13}$ 23 &  E  &  134.9  &  3.46e-04  &  98  &      -          &      -          \\ 
305636.274  & $40_{8,32}$ 39 $\leftarrow$ $39_{9,31}$ 38 &  E  &  333.0  &  3.49e-04  &  158  &  $1.07\pm0.23$  &  $1.32\pm0.19$  \\
305636.275  & $40_{8,32}$ 41 $\leftarrow$ $39_{9,31}$ 40 &  E  &  333.0  &  3.49e-04  &  166  &      -          &      -          \\
305636.295  & $40_{8,32}$ 40 $\leftarrow$ $39_{9,31}$ 39 &  E  &  333.0  &  3.49e-04  &  162  &      -          &      -          \\
305640.828  & $40_{8,32}$ 39 $\leftarrow$ $39_{9,31}$ 38 &  A  &  333.0  &  3.49e-04  &  158  &  $0.92\pm0.11$  &  $1.50\pm0.26$  \\
305640.829  & $40_{8,32}$ 41 $\leftarrow$ $39_{9,31}$ 40 &  A  &  333.0  &  3.49e-04  &  166  &      -          &      -          \\
305640.848  & $40_{8,32}$ 40 $\leftarrow$ $39_{9,31}$ 39 &  A  &  333.0  &  3.49e-04  &  162  &      -          &      -          \\
305861.944  &  $25_{9,16}$ 24 $\leftarrow$ $24_{8,17}$ 23 &  A  &  147.3  &  2.69e-04  &  98  &  $1.21\pm0.13$  &  $1.18\pm0.09$  \\ 
305861.908  &  $25_{9,16}$ 25 $\leftarrow$ $24_{8,17}$ 24 &  A  &  147.3  &  2.69e-04  & 102  &      -          &      -          \\ 
305861.943  &  $25_{9,16}$ 26 $\leftarrow$ $24_{8,17}$ 25 &  A  &  147.3  &  2.70e-04  & 106  &      -          &      -          \\ 
305928.067  &  $25_{9,16}$ 24 $\leftarrow$ $24_{8,17}$ 23 &  E  &  147.3  &  2.23e-04  &  98  &  $0.93\pm0.23$  &  $1.12\pm0.10$  \\ 
305928.036  &  $25_{9,16}$ 25 $\leftarrow$ $24_{8,17}$ 24 &  E  &  147.3  &  2.23e-04  & 102  &      -          &      -          \\ 
305928.066  &  $25_{9,16}$ 26 $\leftarrow$ $24_{8,17}$ 25 &  E  &  147.3  &  2.23e-04  & 106  &      -          &      -          \\ 
\enddata
\tablecomments{
$^{\rm a}$MC contains a quadrupolar nucleus, which is the nitrogen atom ($^{\rm 14}$N), thus having a nuclear spin $I_{\rm N}$ = 1. This leads to an electric interaction, namely the nuclear quadrupole coupling, between the quadrupole moment of nitrogen and the electric field gradient at the nucleus itself. This interaction splits the rotational energy levels and, consequently, rotational transitions, thus giving rise to the so-called hyperfine structure of the rotational spectrum. Therefore, the energy levels involved in each transition are labeled with quantum numbers $J$, $K_{a}$, $K_{c}$, and $F$, where $F$=$J$+$I$. In ALMA observations, these quadrupole hyperfine components were overlapped with each other. $^{\rm b}$Due to the large-amplitude motion of the MC methyl group, each energy level will split into the A-symmetry and E-symmetry species, represented by A and E, respectively.
}
\end{deluxetable*}

\clearpage

\section{$\rm H_{2}$ Column Density from Dust Continuum} \label{sec:appC}

We estimated the molecular hydrogen column density, $N \rm _{H_{2}}$, from the 1 mm dust continuum under the standard assumption of optically thin thermal emission. The beam-averaged column density is derived by the following equation \citep[e.g.,][]{Mar11, Bon19}:

\begin{equation}
N_{{\rm H}_{2}} = \frac{{S_{\rm \nu}}{R_{\rm gd}}}{{\mu}{m_{\rm H}}{\Omega}{{\kappa}_{\nu}}{B_{\nu}}({T_{\rm dust}})}
\label{eq:equation1}
\end{equation}

where $S_{\rm \nu}$ is the integrated continuum flux density within the extraction beam, $R_{\rm gd}$ = 100 is the gas-to-dust mass ratio, $\mu$ = 2.8 is the mean molecular weight per $\rm H_{2}$ molecule \citep{Kau08}, $m_{\rm H}$ is the hydrogen-atom mass, $\Omega$ is the beam solid angle, ${{\kappa}_{\nu}}$ is the dust opacity \citep[adopting ${{\kappa}_{\nu}}$ = 2 cm$^{2}$ g$^{-1}$ at ${\rm \lambda}$${\sim}$1 mm,][]{Bon19} , and ${{B_{\nu}}({T_{\rm dust}})}$ is the Planck function evaluated at the dust temperature $T_{\rm dust}$ = 150 K, following \citet{2020NatAs...4.1170C}. With these assumptions, we obtain $N \rm _{H_{2}}$ = (7.82$\pm$1.56)$\times$10$^{24}$ cm$^{-2}$ for G358.93 MM1, where we conservatively assume a 20\% uncertainty. Molecular abundances relative to $\rm H_{2}$ are then computed as $\chi = N_{\rm t}/N \rm _{H_{2}}$, and the resulting $\chi$ values are reported in Table \ref{table:1}.

\section{Quantum-chemical calculations} \label{sec:appD}

\begin{figure}[ht!]
\centering
\includegraphics[width=0.7\columnwidth]{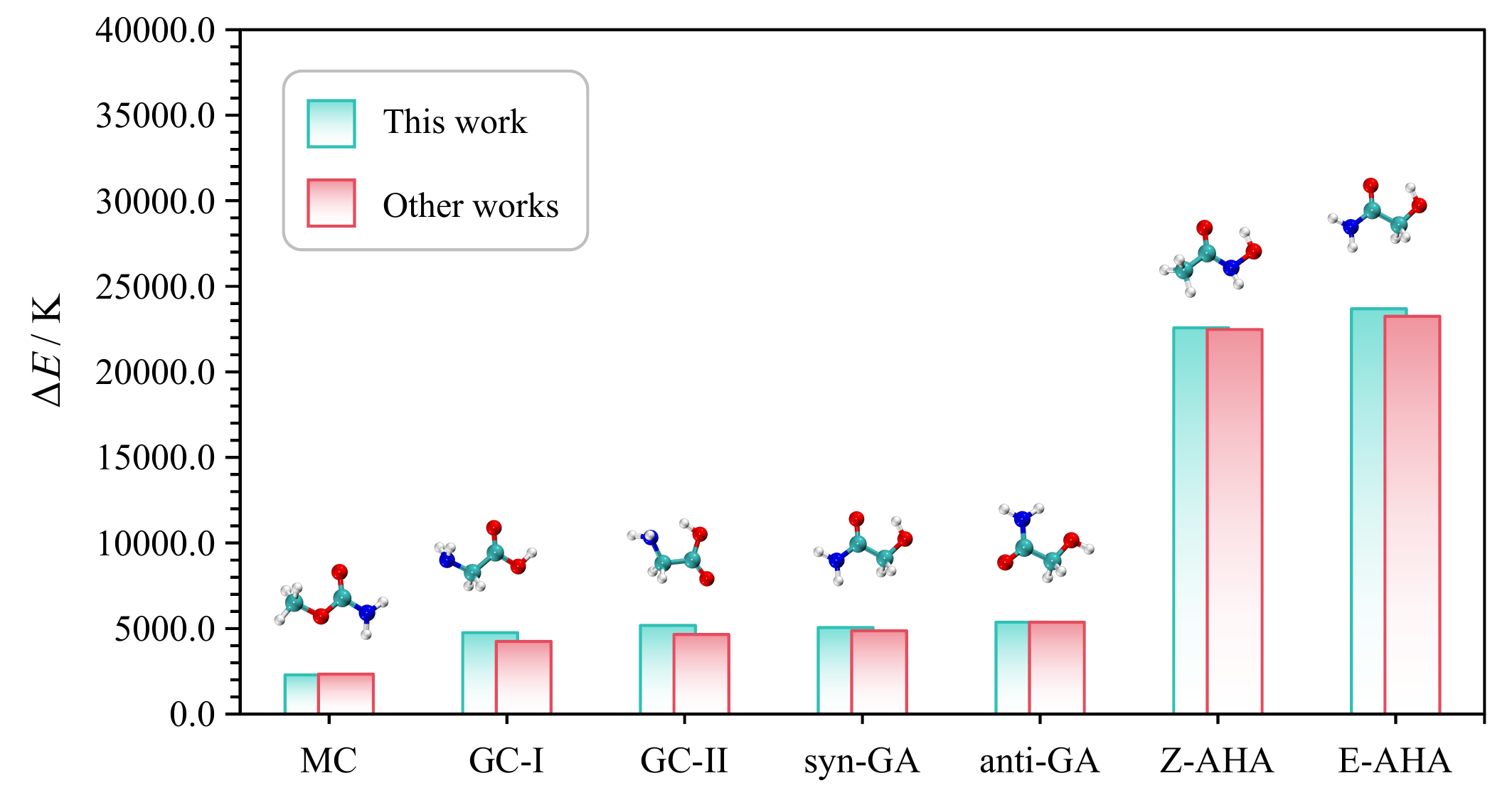}
\caption{Relative energies ($\Delta E$) of spectroscopically accessible C$_{2}$H$_{5}$O$_{2}$N isomers, using the zero-point energies of methylcarbamic acid (CH$_{3}$NHCOOH) as a reference.
\label{fig:5}}
\end{figure}

Although numerous theoretical calculations have been reported for $\rm C_{2}H_{5}O_{2}N$ isomers, they were performed at different levels of theory, making direct comparisons challenging. To place all spectroscopically accessible species considered here on a consistent footing, we re-optimized the geometries of the relevant isomers using previously reported structures as initial inputs and computed their relative energies ($\Delta E$) and dipole moments ($\mu$) at a single level of theory. All calculations were performed with $Gaussian16$ \citep{Gaussian16} at the B2PLYP-D3(BJ)/aug-cc-pVTZ level. Harmonic frequency calculations were used to verify that each optimized structure corresponds to a true minimum (no imaginary frequencies) and obtain zero-point energies (ZPEs). The resulting relative energies and optimized geometries are presented in Figure \ref{fig:5}, consistent with previous theoretical results. A side-by-side comparison with representative literature values is provided in Table \ref{tab:3}.

\begin{deluxetable*}{ccccc}[htp!]
\setlength{\tabcolsep}{30pt}
\tabletypesize{\normalsize}
\tablenum{3}
\tablewidth{0pt}
\tablecaption{Theoretical parameters for all spectroscopically accessible  $\rm C_{2}H_{5}O_{2}N$ isomers \label{tab:3}}
\tablehead{
\colhead{Isomers} & \multicolumn{2}{c}{This work$^{\rm a}$} & \multicolumn{2}{c}{Other work$^{\rm b}$} \\
\colhead{} & \colhead{$\Delta E$ / K} & \colhead{$\mu$ / Debye} & \colhead{$\Delta E$ / K} & \colhead{$\mu$ / Debye}
}
\startdata
 MC  	   &  2288	 & 2.4	&  2335	 & 2.3      \\  
 GC-I	   &  4761	 & 1.2	&  4247	 & 1.1      \\  
 GC-II	 &  5184	 & 5.5	&  4650	 & 5.5      \\  
 syn-GA	 &  5055	 & 4.4	&  4866	 & 4.4      \\  
 anti-GA &  5375	 & 3.5	&  5367	 & 2.6      \\  
 Z-AHA	 &  22572	 & 3.3	&  22474	&  3.3    \\  
 E-AHA	 &  23684	 & 3.2	&  23247	&  3.2    \\  
\enddata
\tablecomments{
$^{\rm a}$Calculated at the B2PLYP-D3(BJ)/aug-cc-pVTZ level in this work.
$^{\rm b}$Theoretical data are from \citet{1995ApJ...455L.201L, Maris04, 2011AA...532A..39L, 2019ESC.....3.1170S, 2020AA...639A.135S, 2022AA...666A.134S}.
}
\end{deluxetable*}

\section{Identification of CH$_{3}$OH and $^{13}$CH$_{3}$OH} \label{sec:appE}

\begin{figure}[ht!]
\centering
\includegraphics[width=1\columnwidth]{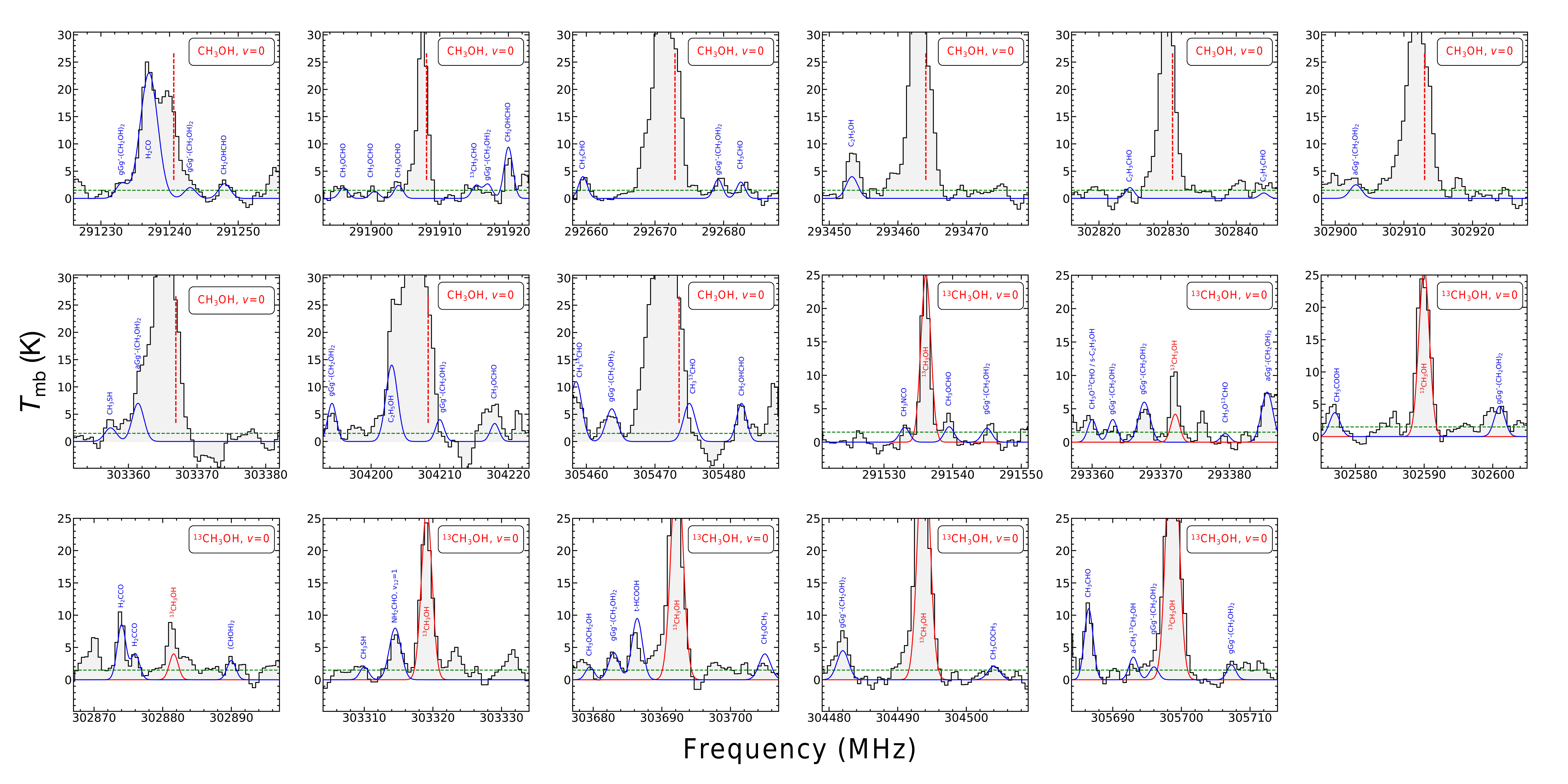}
\caption{Observed (black) and LTE modeled (red) spectra of $^{13}$CH$_{3}$OH toward G358.93 MM1, together with the corresponding CH$_{3}$OH transitions used for line identification. No LTE modeled spectra are shown for CH$_{3}$OH because the main-isotopologue lines are optically thick. Vertical dashed lines mark the rest frequencies of the CH$_{3}$OH transitions. All other plotting conventions follow Figure \ref{fig:1}.
\label{fig:6}}
\end{figure}

To quantify the MC-to-methanol ratio in G358.93 MM1, we searched for transitions of methanol (CH$_{3}$OH) and its $^{13}$C isotopologue ($^{13}$CH$_{3}$OH), as shown in Figure \ref{fig:6}. Because the main-isotopologue CH$_{3}$OH lines are optically thick in this dataset, we derived the methanol column density from the optically thin $^{13}$CH$_{3}$OH emission and then scaled by the $^{12}$C/$^{13}$C isotopic ratio. The $^{12}$C/$^{13}$C ratio was computed following \citet{2023A&A...670A..98Y}. We obtain $N_{\rm ^{13}CH_{3}OH}$ = (7.79$\pm$1.56)$\times$10$^{16}$ cm$^{-2}$, which corresponds to $N_{\rm CH_{3}OH}$ = (2.16$\pm$0.57)$\times$10$^{18}$ cm$^{-2}$.

\bibliography{MC}{}
\bibliographystyle{aasjournalv7}

\end{document}